\documentclass[conference]{IEEEtran}
\IEEEoverridecommandlockouts
\usepackage{cite}
\usepackage{amsmath,amssymb,amsfonts}
\usepackage{algorithmic}
\usepackage{graphicx}
\usepackage{subcaption}
\usepackage{textcomp}
\usepackage{xcolor}
\usepackage{diagbox}
\usepackage[shortlabels]{enumitem}
\usepackage{tikz}

  \newcommand\copyrightbeforepulbishing{
  This version of the manuscript is the original author-submitted version to IEEE International Symposium on Technology and Society (ISTAS 2025). Please cite final version available on IEEE Xplore as  S. Dulam and C. L. Dancy, "Computational Cognitive Modeling to understand the effects of Racializing AI on Human-AI cooperation with PigChase Task," 2025 IEEE International Symposium on Technology and Society (ISTAS), Santa Clara, CA, USA, 2025, pp. 1-9, doi: 10.1109/ISTAS65609.2025.11269626.
  }
\newcommand\copyrightnotice{%
\begin{tikzpicture}[remember picture,overlay]
\node[anchor=south,yshift=10pt] at (current page.south) 
  {\fbox{\parbox{\dimexpr\textwidth-\fboxsep-\fboxrule\relax}{\copyrightbeforepulbishing}}};
\end{tikzpicture}%
}
\def\BibTeX{{\rm B\kern-.05em{\sc i\kern-.025em b}\kern-.08em
    T\kern-.1667em\lower.7ex\hbox{E}\kern-.125emX}}
\begin{document}
\title{Computational Cognitive Modeling to understand the effects of Racializing AI on Human-AI cooperation with PigChase Task\\
\thanks{This work was supported by the National Science Foundation under grant No. 2144887}
\author{\IEEEauthorblockN{Swapnika Dulam}
\IEEEauthorblockA{\textit{Department of Computer Science and Engineering} \\
\textit{Pennsylvania State University}\\
PA, USA \\
szd5775@psu.edu}
\and
\IEEEauthorblockN{Christopher L. Dancy}
\IEEEauthorblockA{ {Department of Industrial and Manufacturing Engineering }\\
\textit{Pennsylvania State University}\\
PA, USA \\
cdancy@psu.edu}
}
}
\maketitle
\copyrightnotice
\begin{abstract}
Despite the continued anthropomorphization of AI systems, the potential impact of racialization during human-AI interaction is understudied. This study explores how human-AI cooperation may be impacted by the belief that data used to train an AI system is racialized, that is, it was trained on data from a specific group of people.  
During this study, participants completed a human-AI cooperation task using the Pig Chase game, a variant of the Stag Hunt game. Participants of different self-identified demographics interacted with AI agents whose perceived racial identities were manipulated, allowing us to assess how sociocultural perspectives influence the decision-making of participants in the game. After the game, participants completed a survey questionnaire to explain the strategies they used while playing the game and to understand the perceived intelligence of their AI teammates. 
Statistical analysis of task behavior data revealed a statistically significant effect of the participant's demographic, as well as the interaction between this self-identified demographic and the treatment condition (i.e., the perceived demographic of the agent). The results indicated that Non-White participants viewed AI agents racialized as White in a positive way compared to AI agents racialized as Black. Both Black and White participants viewed the AI agent in the control treatment in a negative way. 
A baseline cognitive model of the task using ACT-R cognitive architecture was used to understand a cognitive-level, process-based explanation of the participants’ perspectives based on results found from the study. This model helps us better understand the factors affecting the decision-making strategies of the game participants. Results from analysis of these data, as well as cognitive modeling, indicate a need to expand understanding of the ways racialization (whether implicit or explicit) impacts interaction with AI systems.
\end{abstract}


\section{Introduction}
With the increase in the proliferation of AI-based systems in day-to-day activities, understanding how the racialization of AI systems may impact the way people interact with them is important, given both the ubiquity of those systems and the sustained impact of systems of racism on societies. The Computers Are Social Actors (CASA) \cite{nass1994a} paradigm describes how humans unthinkably apply the same social heuristics used for human interactions to interactions with computers. Nass et al. \cite{nass1994a} described a study where human participants displayed more human-like interactions with computers after the researchers strategically manipulated visual or verbal cues. Further, ethnicity related cues have triggered social responses similar to those observed in human interactions \cite{nass2000machines}. AI-based systems such as Apple's Siri, Amazon's Alexa, and OpenAI's ChatGPT become increasingly integrated into daily life, it is important to address concerns arising in such interactions \cite{abercrombie2021a}, especially from a socio-cultural context. 
Despite the continued anthropomorphization of AI systems, the potential impact of racialization is understudied, but nonetheless important given the push for humans collaborating with AI agents in
team-like settings to accomplish tasks, from disaster relief \cite{moitra2022a} to assisted care \cite{schicktanz2023a}. 

There have been several cases where the AI systems trained on various datasets amplify the existing biases of society and reinforce its stereotypes; specifically, systems used for decision-making that result in hiring decisions favoring certain races and genders \cite{dastin2025a}, restricted Amazon’s same-day deliveries to specific neighborhoods \cite{ingold2025a}, higher mortgage \cite{bartlett2022a}, or medical costs \cite{ferrara2025a}, as well as several known issues with recidivism focused systems like COMPAS \cite{flores2016a}. These studies demonstrated how AI-systems can encode existing sociocultural structures, thereby harming minoritized individuals and communities. These structures are encoded in the AI artifacts from the datasets themselves or during stages of development or deployment \cite{livingston2020a}, \cite{dancy2022a}, \cite{workman2023identifying}. Therefore, ethical choices in AI development have seen an increase in research studies. While most of these studies analyze the datasets or change specific parameters in these systems to mitigate their biases  \cite{hort2024a}, there has been limited research on how human behaviors or their implicit biases affect AI artifact development and maintenance, which might ultimately affect its users. Therefore, it would be helpful for an AI system developer to have a better understanding of the humans affected directly or indirectly by AI systems and the bidirectional interaction of AI systems with humans through a sociocultural lens.

Just as race influences human-to-human interactions, it also impacts how people perceive, trust, and collaborate with AI agents \cite{nass2000machines}. These challenges in human-AI interactions become even more nuanced when considering how the race and culture of the individual teammates shape their interactions with the AI agent in a human-AI team. Racialized perceptions of AI—whether explicit or implicit—can affect trust, satisfaction, and the equitable distribution of influence in teams; several such cases have been demonstrated in studies highlighting shooter bias (e.g., see \cite{correll2002a}], \cite{bartneck2018a}, where the White participants were quick to shoot the Black targets). Davis et al. \cite{davis2023a} have also shown that racial stereotypes impact how consumers interact with anthropomorphic AI agents, with different racialized AI bots being perceived and treated based on existing stereotypes. In a study conducted to investigate whether people trust partners from different racial backgrounds, Stanley et al. \cite{stanley2011a} also found that participants placed higher trust in partners who shared the same self-identified racial background. 

To develop AI systems that are socioculturally competent, we need to understand “How humans view these AI systems through a racial lens?” and “How do AI systems treat people of different races?”. Understanding the former question can help us develop more ethical AI systems. In this paper, we look at a human-AI interaction experiment, where people interact with an AI artifact in a racialized setting to understand their behaviors toward racialized AI agents. 

To model cooperation in human-AI teaming efforts, we turn to the classic Prisoner's Dilemma  \cite{lebiere2000a}, which has significantly influenced research on cooperation and social well-being. A variation of this problem that is inspired by nature, known as the Stag Hunt task, was introduced by Skyrms \cite{skyrms2003a}. In the Stag Hunt task, two hunters can either individually catch rabbits, giving a small but guaranteed reward, or join forces to \textquotedblleft hunt\textquotedblright \hspace{0.2em} a stag, providing enough food for several days but requiring mutual cooperation.  Yoshida et al. developed a \textquotedblleft Game Theory of Mind\textquotedblright \hspace{0.2em} approach for the Stag Hunt task \cite{yoshida2008a}, further exploring the strategic decision-making involved in cooperative interactions. The Microsoft Malmo Collaborative AI Challenge \cite{johnson2016a} created a Minecraft-based variation of this stag hunt task and named it Pig Chase. This AI challenge invited participants to develop an AI agent that could score higher by effectively collaborating with any collaborator. Atkins et al. \cite{atkins2021a} developed a simplified JavaScript version of the Pig Chase task. Atkins et al. \cite{atkins2021a} showed how the racialization of an AI agent could affect behavior during a human-AI cooperation task. In that study, the participants were told that the AI agent they would be interacting with was trained by observing the behavior of certain races, indicating that they interacted with a racialized AI agent. The results showed that the self-identified race interacted with how the AI agent was racialized to ultimately impact their performance while cooperating with the AI agent during the Pig Chase game. Here, we describe the results from an expanded version of that study that included participants from a wider range of self-identified racial categories and treatment conditions that included pictures of racialized individuals (which presumably added a more explicit visual phenotypical racial association to the AI agent which was not present in \cite{atkins2021a}). 

Statistical analysis of the data collected from the task gives insights into participants' strategies when they played the Pig Chase game. This analysis, however, can only uncover what strategies participants may have used in the Pig Chase game and cannot tell us more about potential cognitive processes undergirding strategy use, that is, mechanistic explanations \cite{thagard2018social}. However, since our interest is in understanding the potential impact of racialization on these strategies, a model is needed that can uncover the cognitive processes behind such decisions. We use the Adaptive Control of Thought - Rational (ACT-R) cognitive architecture \cite{anderson2007a} to model participant strategies in the game. 

Computational cognitive modeling can be used to simulate cognitive processes in decision-making scenarios to help understand and predict human behavior (eg. \cite{west2006a}, \cite{ritter2019a}. ACT-R is an open-source cognitive architecture that has an active development community and is widely used to model cognition. Several ACT-R models have been built to understand the strategy of the participants and to formulate a theory about behaviors \cite{ritter2019a}, \cite{anderson2019a}, \cite{anderson2021a}. We use an ACT-R model to understand dominant strategies used by participants from various demographics and their interactions with racialized AI agents in their respective treatment conditions. By examining these factors, we aim to contribute to developing AI systems that not only improve performance but promote fairness, inclusion, and equity in increasingly diverse human-AI teams.

In this paper, we aim to answer the following questions:
(1) What strategies did participants use when they interacted with AI agents during the Pig Chase game? 
(2) How might the racialization of those AI agents impact strategies used during the task? 
In the next section, we detail the Pig Chase experiment and present the results from the collected data, followed by the ACT-R cognitive model that helps us understand participant strategies and discuss them.

\section{Pig Chase Experiment}
The Pig Chase experiment data collection happened through an online platform. In this section, we explain the experiment procedure in detail.
\subsection{Participants }
The participant sample included $1008$ participants, all recruited through Prolific.co. Participants were paid \$$7.50$ for their participation, which was estimated to finish in $45$ minutes. Prolific’s automatic participant filters were used to specify participants’ demographics for their self-reported race and being from and located within the United States to achieve a balanced set of people who identified as “Black/African American” (Black for short), “White/ Caucasian” (White for short), or [“Asian”, “Mixed”, or “Other”] grouped into a “Non-White” category. 
\subsection{Materials }
The participants were asked to play the Pig Chase game with an AI agent for fifteen trials. The participants could either trust their companion, an AI agent, and decide to “catch the pig” together for a higher reward of $25$ points or exit the game through designated “exit blocks” for a lower reward of $5$ points.

Each participant was placed into one of seven possible treatment conditions, where they were told that “the AI learned by observing the behavior of people who identify with a certain racial category,” or a control condition that did not mention race. The treatment conditions were:  
1. \textit{B1},\textit{ B2}: Black or African American, with one of two possible pictures displayed for 
reference (these and other pictures were obtained from the Chicago Face Database \cite{ma2015a} and a 
different one was used for each treatment condition). 
2. \textit{BNP}: Black or African American, but with no picture shown. 
3. \textit{W1},\textit{ W2}: White or Caucasian, and with one of two possible pictures displayed for 
reference. 
4. \textit{WNP}: White or Caucasian, but with no picture was shown. 
5. \textit{Control}: Did not mention race and was told that AI was trained by observing people’s 
behaviors without any race-specific details or pictures. 
The AI agent with which participants were teamed up used a Wizard-of-Oz technique \cite{dahlback1993wizard} and 
was not trained by observing human behavior. Instead, they all used an A* algorithm to catch the pig. A* is a simple path-finding algorithm that enables the AI agent to find the shortest path towards the pig. Individual task-related behaviors, such as keys pressed, reaction times, and scores, were collected for each round during the task. 
\subsection{Procedure}
The participants started on the Prolific.co website and were first directed to the Qualtrics page with game instructions. Next, they were directed to Pavlovia.org, where they played the game. Finally, they were again redirected to Qualtrics for the post-game survey. These steps are discussed below in detail.
\subsubsection{Pre-game steps}
Participants began the experiment through Prolific.co website, where they were randomly assigned treatment conditions. They were then directed to a Qualtrics page with specific instructions for the game and details specific to their treatment conditions. They were informed that the first three trials in the study were for practice to help them understand the game. Additionally, they were instructed to exit through the rightmost exit on the eighth trial to ensure they were paying attention. After reading these instructions, participants were finally redirected to play the game hosted through Pavlovia.org. 

\subsubsection{Pig Chase Game}
\begin{figure}[!ht]
\begin{subfigure}[h]{0.45\linewidth}
\includegraphics[width=\linewidth]{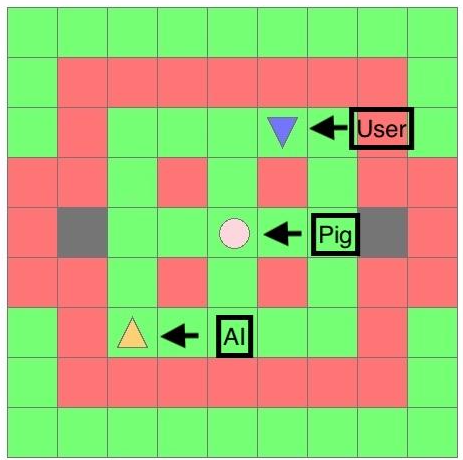}
\caption{Start game position}
\end{subfigure}
\hfill
\begin{subfigure}[h]{0.45\linewidth}
\includegraphics[width=\linewidth]{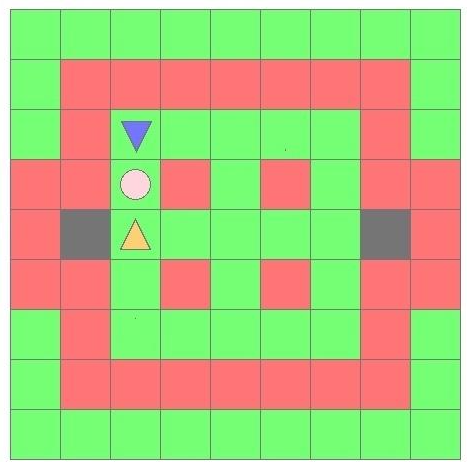}
\caption{One end game position}
\end{subfigure}%
\caption{Pig Chase Game}
\label{1}
\end{figure}

The participants interacted with the interface, as shown in Fig.\ref{1} (a), where they controlled a blue, triangular game piece and collaborated with an “AI-controlled” yellow, triangular game piece to catch a pink, circular game piece representing a pig. The direction of the triangular game pieces indicated their orientation. The game used simple controls, and the arrow keys moved the game piece in that direction. Therefore, no prior game-playing experience was needed. The first three were the learning trials in which the participants could learn about controls and form strategies. All pieces were on a $9$ x $9$ grid but could only move within a $5$ x $5$ area (green tiles), which was blocked by red square tiles. Moreover, each action in the game deducted $1$ point from the score, and the participants were supposed to catch the pig before they exhausted the limit of $25$ actions per trial.

\subsubsection{Post-Game Survey}
After participants completed all $15$ trials, they were redirected from Pavlovia to Qualtrics to answer five qualitative questions about their actions in the experiment. The questions were designed to gain a deeper understanding of participants' strategies as they played the game and their perception of the AI agent.

\section{Statistical analysis}
The data collected from $1008$ participants was later cleaned for further analysis. First, we removed incomplete records and were left with $950$ records. These records were further filtered for duplicate IDs (indicating multiple attempts by the same participant), leaving data from $939$ participants. We then removed the outliers for every treatment by calculating the \textit{z-score} and removing records based on the score field where the absolute value of the \textit{z-score} was greater than $3$. This removed $4$ records from the $939$ records, resulting in $935$ records. Table \ref{t1} below shows the participant split for all demographics and treatment conditions. 

The analysis in the following sections was performed at two levels, the treatment level and the treatment group level. At the treatment level, there are $7$ treatment conditions: \textit{B1, B2, BNP, W1, W2, WNP,} and \textit{Control} treatments. These individual treatment conditions were grouped such that\textit{ B1, B2,} and \textit{BNP} are grouped as the \textit{Black} treatment group, and\textit{ W1, W2,} and \textit{WNP} are grouped as the \textit{White} treatment group, and the same \textit{Control} treatment group from the previous level was used. The individual treatment-wise comparisons have been done wherever possible, but sometimes, if there was a similar trend in the group, we discuss the analysis on the treatment group to simplify the discussion.
\begin{table}[htbp]
\caption{Participant demographics for each treatment condition}
\begin{center}
\begin{tabular}{lcccc}
\hline
Treatment & Group & Black & White & Non-White \\
\hline
\textit{B1}        & \textit{Black} & 47    & 48    & 48        \\
\textit{B2 }       & \textit{Black} & 46    & 47    & 44        \\
\textit{BNP}       & \textit{Black} & 46    & 44    & 43        \\
\textit{Control}   & \textit{Control} & 42    & 48    & 43        \\
\textit{W1}        & \textit{White} & 44    & 48    & 44        \\
\textit{W2}        & \textit{White} & 47    & 45    & 32        \\
\textit{WNP}       & \textit{White} & 42    & 47    & 40      \\
\hline
\end{tabular}
\end{center}
\label{t1}
\end{table}

\subsection{ Quantitative Analysis }
We compared participant scores in various treatment conditions, performed an ANOVA analysis, and correlated participant scores with their intelligent estimates of the AI agent they interacted with. We excluded these records from the first $3$ trials as they were learning trials. 

\subsubsection{Analysis of Participant Scores}
\begin{figure}[!ht]
    \centering
    \includegraphics[width=\linewidth]{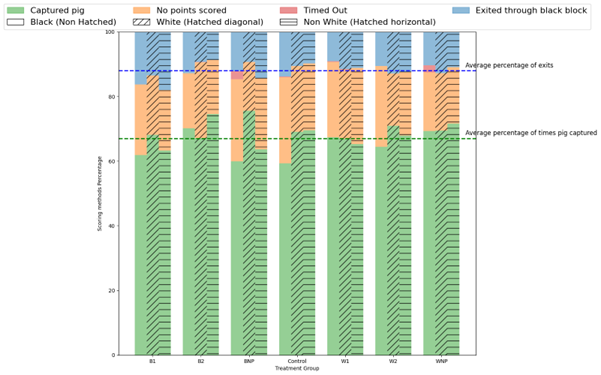}
    \caption{Percentage of scoring methods used by participants in all treatment conditions}
    \label{f2}
\end{figure}

$935$ participants scored points in trials $67$\% of the time, exited through the designated block only $12$\% of the time, exhausted their actions $21$\% of the time, and timed out only $1$\% of the time, indicating the game was played to win by cooperating with AI agent. This remained consistent with all treatment conditions, as shown in Fig. \ref{f2} where the participants in all treatment groups captured the pig $60$-$70$\% of the time, chose to exit $9$-$12$\% of the time, and did not score in the remaining trials. The average percentage of catching the pig and exiting for all treatment conditions are indicated by respective colored lines to enable easy comparison. Although catching the pig was consistent across all treatment conditions, the number of actions taken to catch the pig affected the scores secured by the players. A higher number of actions taken by the participants resulted in a lower score due to the penalty incurred for every action.

\subsubsection{ANOVA}
For the $935$ records, as displayed in Table \ref{t1}, a two-way ANOVA was used to examine the effect of the treatment condition and participant demographics on the cumulative scores. The $3$ x $7$ (participant demographic x treatment condition) ANOVA of participants’ final scores revealed a significant effect of participant demographic ($F = 6.85, p < .001$) and treatment X demographic ($F = 2.22, p < .01$) but not treatment ($F =1.66, p = .12$). These results indicate that participants’ demographics significantly influenced the obtained scores, with the interaction suggesting that the effect of treatment varied across demographic groups.

\subsubsection{Correlation of participant scores with perceived intelligence of the AI agent}
\begin{figure}[!ht]
    \centering
    \includegraphics[width=\linewidth]{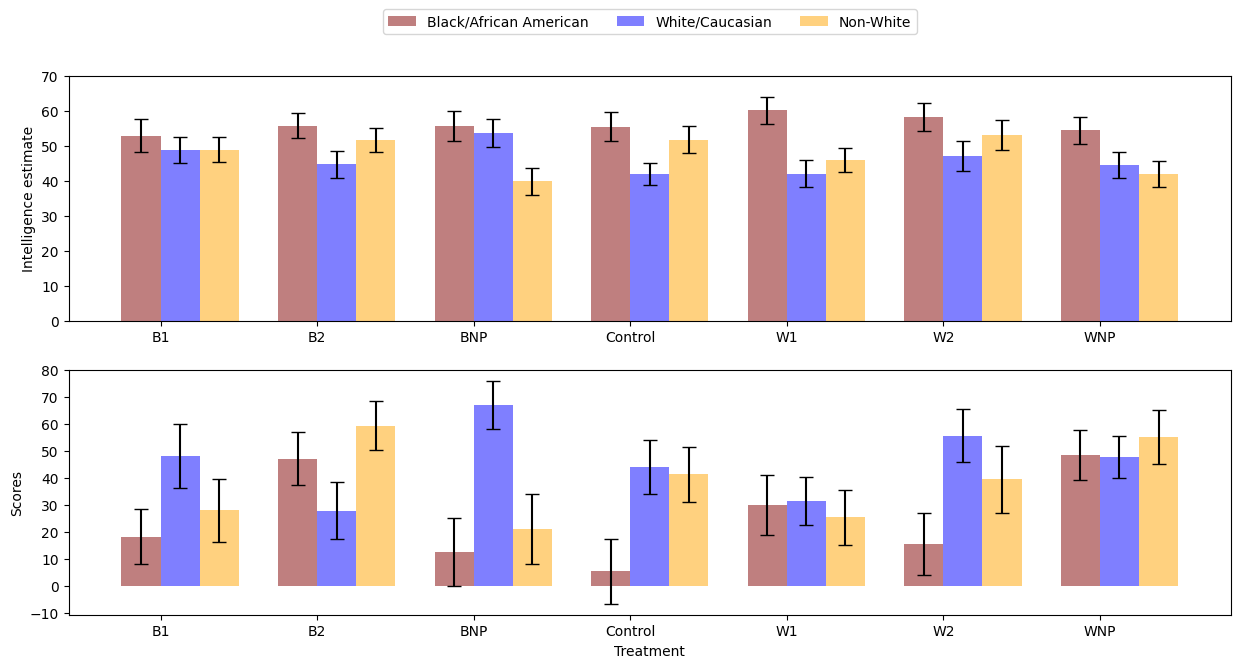}
    \caption{Average scores and intelligence estimates for all treatment conditions}
    \label{f4}
\end{figure}

 One of the post-game survey questions asked the participants to rate the intelligence of the AI agent they interacted with. The participants used a slider from $0$ to $100$ to indicate their intelligence estimate. We analyzed the results to see if there was a correlation between the scores and this perceived intelligence. We hypothesized that a higher score for a participant meant that the participant coordinated with the AI agent efficiently to catch the pig. In this case, they would believe the AI agent to be helpful and rate higher intelligence for the AI agent. Fig. \ref{f4} shows average intelligence estimations by participants of all treatments and demographics on top, average scores by participants, and mean scores by all treatments and demographics at the bottom. However, this correlation is not seen in participants of the Black demographic, who rated the AI agent with higher intelligence estimates, irrespective of the score. To verify the hypothesis, we checked the Pearson correlation of the total score obtained by participants with their corresponding intelligence estimates. We observe a positive correlation with a significant effect for White participants in the \textit{B1} treatment condition ($r > 0.5$) and a weaker correlation in the \textit{B2} treatment condition ($r > 0.3$), similarly, weaker correlation for Non-White participants in the \textit{WNP} ($r > 0.4$), \textit{W2} ($r > 0.3$), and \textit{BNP} ($r > 0.3$) treatment conditions. 

\subsection{Qualitative analysis}
We categorized the data into seven labels based on labeling reported by Atkins et al. \cite{atkins2021a} \textquotedblleft AI Cooperated with Human\textquotedblright  \hspace{0.2em} categorized all responses where the participants had a positive opinion on the AI agent. The labels \textquotedblleft AI Movement was User Dependent\textquotedblright, \textquotedblleft AI had no pattern\textquotedblright, \textquotedblleft Vague\textquotedblright, \textquotedblleft User Focused on own Movement\textquotedblright \hspace{0.2em} are labels where the participants' opinions towards AI are neutral. While the labels \textquotedblleft AI not intelligent\textquotedblright \hspace{0.2em} and \textquotedblleft AI Worked against Human\textquotedblright \hspace{0.2em} indicate a negative opinion towards the AI agent.

Two researchers coded the data from survey responses to classify the labels. This data was later combined into a single set of responses. The score of the weighted Quadratic Kappa was $0.73$, indicating substantial agreement between our coded responses. We chose weighted Quadratic Kappa because our labels are interrelated, and the mismatches will have different penalties. 

\begin{figure} [!ht]
    \centering
    \includegraphics[width=\linewidth]{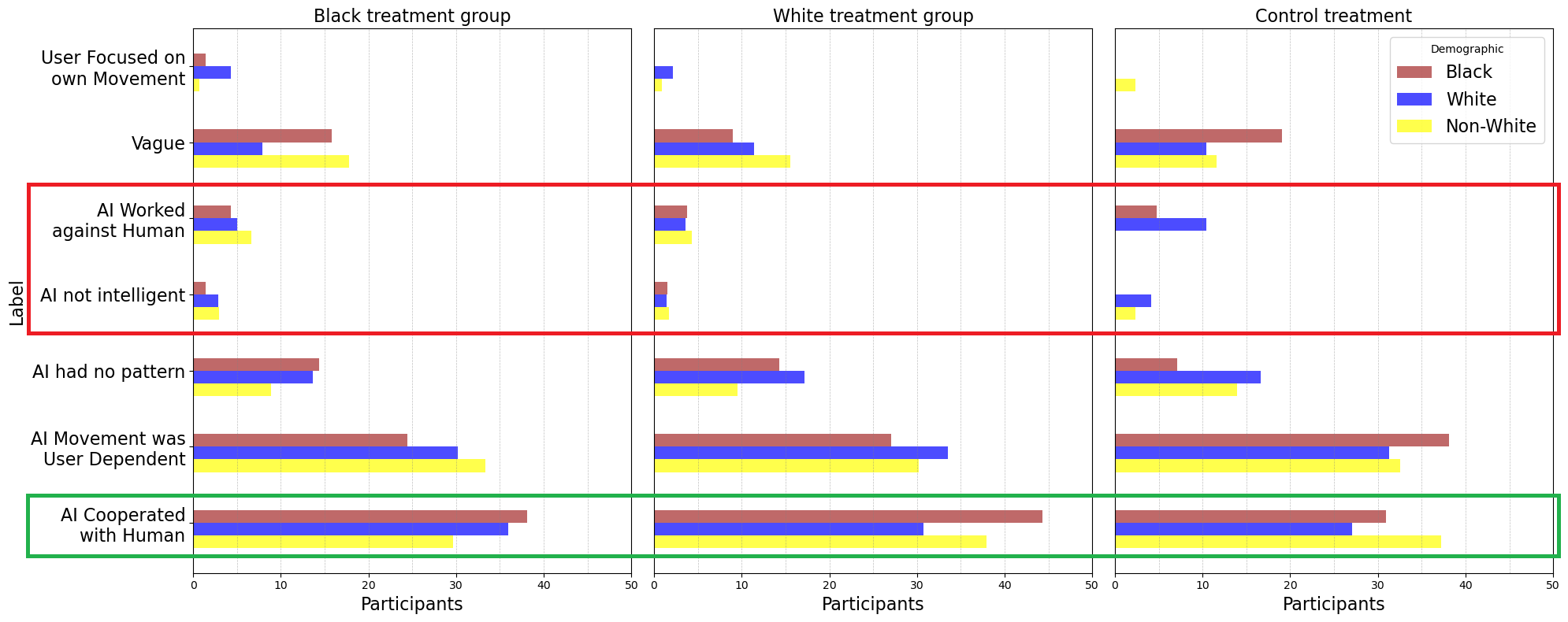}
    \caption{Participant responses in grouped treatment conditions (in percentages).}
    \label{f5}
\end{figure}

To depict the overall trend, we represent the responses at the grouped treatment level. Fig. \ref{f5} shows the coded responses for all participants in grouped treatment conditions (\textit{Black, White, Control}). The important differences for some labels are highlighted with a rectangular box. Here, we discuss the trends in these positive and negative opinions.

\textbf{“AI cooperated with Human”}: Black participants believed the AI agent cooperated with them the most in \textit{White} and \textit{Black} treatment conditions. White participants reported that the AI agent cooperated marginally better in \textit{Black} treatment conditions, followed by \textit{White} and \textit{Control} treatment conditions. Non-White participants believed the AI agent cooperated the least in \textit{Black} treatment conditions. 

\textbf{“AI not intelligent”}: Again, compared to other demographics, fewer Black participants believed the AI agent was not intelligent. While the White and Non-White participants believed the AI agent in the \textit{Black} and \textit{Control} treatment conditions was not intelligent.

\textbf{“AI worked against Human”}: The trends mostly remain same as the label \textquotedblleft AI not intelligent\textquotedblright, with subtle variations in the \textit{Control} treatment where higher number of White participants believed the AI agent worked against them, and not a single Non-White participant believed the AI agent worked against them in the \textit{Control} treatment.

\subsection{Summary of statistical analysis}
 The Pig Chase game has simple instructions that do not require prior game-playing experience, but it is strategy-heavy. This means that a player needs to analyze the strategy of an AI agent to optimize their score. They can choose either a risk-dominant strategy by proceeding to the nearest exit and getting 5 points, or pick a payoff-dominant strategy to catch the pig, which can be strategically executed to gain higher points, but also might result in not being able to capture the pig sometimes. In the payoff-dominant strategy, the players trust the AI agent and cooperate with it to catch the pig. As shown in Fig. \ref{f2}, the number of exits indicates a selfish strategy or a lack of cooperation with the AI agent. This was marginally higher in the \textit{Black} treatment conditions compared to the \textit{White} treatment conditions, among players of the Non-White demographic, followed by participants of the Black demographic. This is reflected in their lower scores obtained in \textit{B1} and \textit{BNP} treatment conditions by Black and Non-White participants. 

Despite receiving lower scores, the Black participants rated the AI agent as having higher intelligence. The intelligent estimates for other demographics correlated with their scores, specifically for the White participants in the \textit{Black} treatment conditions and Non-White participants in the \textit{White}, \textit{Black} treatment conditions, as can be seen in Fig. \ref{f4}. Qualitative analysis of survey answers revealed that Black participants believed that AI cooperated in all treatment conditions (see Fig. \ref{f5}), as was also visible in intelligence estimates (see Fig. \ref{f4}). White participants had negative opinions of AI in the \textit{Control}, \textit{Black} treatment conditions compared to \textit{White} treatment conditions. Non-White participants believed AI cooperated less and worked against them more in \textit{Black} treatment conditions compared to \textit{White} or \textit{Control} treatment conditions. All the demographics had the highest positive opinion and the least negative opinion of the AI agent in the \textit{White} treatment conditions.

\section{ACT-R cognitive modeling}

Cognitive modeling through cognitive architectures can be used to better understand the cognitive processes involved in these decision-making strategies. Cognitive architecture as described by Anderson in \cite{anderson2007a} is, \textquotedblleft Cognitive architecture specifies the structure of the brain at a level of abstraction that explains how it achieves the function of the mind\textquotedblright. It proposes artificial computational processes that act like human cognitive processes, thereby explaining cognition using psychological theory. 

Many cognitive models are computer programs written in a programming language that run on a particular cognitive architecture. Such cognitive models provide an approximation of cognition to simulate human behavior\cite{ritter2014a}. A distinctive feature of these cognitive models from other computing models is the ability to account for limitations of human abilities. 

The ACT-R cognitive architecture \cite{anderson2007a} defines basic cognitive and perceptual operations that enable the human mind. The primary implementation of ACT-R architecture is in Common Lisp, which we used to build this model as well. ACT-R contains a collection of independent computational systems called modules that operate in parallel but are constrained by a central production system (procedural memory). These modules run asynchronously. 

ACT-R has evolved over time, with more recent versions having eight main modules and a central production system. These eight modules can be further divided into internal and external Modules. The external modules are input modules, including vision, audio, output, motor, and speech. These modules serve as a means of communication between the ACT-R rule-based engine, called the production system, and the external environment. The internal modules include imaginal, temporal, goal, and declarative Modules. These modules are used to internally drive various state representations and communicate with the production system, which represents the procedural memory system. This communication is specifically achieved with the help of buffers associated with each module. The production system is a type of rule pattern-matching system that checks various buffers of respective modules to select the next rule based on the utility computed by the utility module. 

 The production system automatically keeps track of the utility associated with each rule to choose a rule with better utility in the case of conflicts between multiple rules that can be selected for the given scenario.
 The learning utilities for a production \(i\) after its \(n^{th}\) usage is defined as:
\[ U_i (n)=U_i (n-1)+ \alpha [R_i (n)-U_i (n-1)] \]
where \(\alpha\) is the learning rate set by the \textit{:alpha} parameter with a default value of $0.2$; \( R_i (n) \) is the effective reward value given to production \(i\) for the \(n^{th}\)usage. 

\subsection{The Pig Chase Cognitive Model}
The Pig Chase game for the study was coded using JavaScript and socket communication to interact with the ACT-R server on a designated port. The game positions of both the agents and the pig were sent to the ACT-R model using a visual location buffer (which keeps track of visual locations). The ACT-R model simulates a participant's behavior to move the blue triangular game piece (see Fig. \ref{1}) to capture the pig along with the AI agent (yellow triangular game piece).

 The ACT-R model uses the production system to consider the current game position and possible movements, ultimately using imaginal, retrieval, and goal buffers to decide the next movement. 
 
\subsection{Modeling strategies }

When the game begins, the model's goal state is set to find the pig. It first checks the Exit Strategy to see if it decides to catch the pig or exit. Then, the model checks if it is blocked and determines the cause of the block so that appropriate action can be taken. If the pig caused the block, the agent could proceed to catch the pig, subject to collaboration with the AI agent. Later, the model decides the next best action, to be in a better spot to catch the pig. The utilities of the production rules are set to mimic the game rules by rewarding catching the pig with 25, choosing an exit with 5, and penalizing every action with $-1$. Various strategies used to make decisions by the model are described below in detail.

\begin{itemize}
    \item Exit Strategy: The ACT-R agent’s exit strategy is based on the AI agent’s proximity to the pig. The proximity is calculated based on the Manhattan distance between the AI agent and the pig. If the AI agent’s proximity towards the pig increases, the ACT-R agent will proceed to capture the pig, or it will check once more the proximity of the AI agent to the pig, and if it is still increasing, it will exit.
     \item Check blocked: After determining to proceed, the rules next check if the agent is blocked. The agent can be blocked by a pig, an AI agent, or a red block. If the agent is blocked by the pig, it implies that the pig is found. The next step would be to see if the AI agent is near and use the rotation strategy until the AI agent blocks the pig from the other end. However, if the ACT-R agent is blocked by the AI agent or a red block, the model checks to see if it cannot move and randomly picks a different direction to move. 
     \item  Rotation Strategy: The Pig Chase game is played in a turn-based setup, meaning the AI agent, which is based on the A* algorithm, moves only when the user or ACT-R agent moves. However, many participants may have failed to figure out this strategy, leading to lower scores. To model this behavior, the ACT-R agent had a rotating strategy encoded with a negative reward mechanism. The agent rotated anticlockwise in the same location while standing adjacent to the pig and waiting for the AI agent to block the pig from the other end. The model more closely mimicked the participant's behavior by reducing the activation of rotation-related chunks. This was achieved by setting the base-level activation parameter to $-0.15$ (after trying $-0.3$ and $-0.2$, which did not lead to a closer fit). This value can be changed further to accurately represent data in each treatment condition for various participant demographics.
    \item Navigation strategy: If the ACT-R agent has enough actions left ($>20$) and is not blocked, it plans its next move to capture the pig. To accomplish this, two things must be considered: possible actions from the current location and the pig's location. The possible actions from the current location are static for any given game location and therefore are encoded as chunks in declarative memory to be retrieved at each stage. Further, the pig's location is also considered when choosing the next best action based on orientation. Changing the orientation requires another action and hence is penalized with $-1$ according to game rules. So, there is a trade-off between choosing fewer actions and choosing the best possible location, which is determined by the probability of production rule selection. These probabilities are updated by the utility learning module discussed earlier. After the action is decided, the motor module presses the required key. The loop is repeated to check the actions taken and choose between an exit strategy and deciding to catch the pig until the pig is caught.

\end{itemize}

\subsection{Model performance}
\begin{figure} [!ht]
    \centering
    \includegraphics[width=\linewidth]{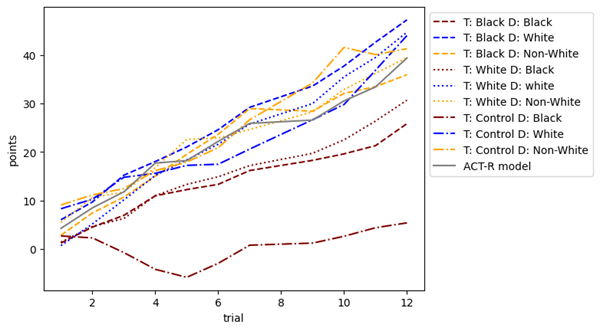}
    \caption{Average cumulative score per trial for ACT-R model compared with data from all grouped treatment conditions}
    \label{f8}
\end{figure}

The model was run over $150$ times to collect the data, and each run took $20-25$ minutes to finish. The results of the average cumulative score per trial are displayed in Fig. \ref{f8}.
 To see the model fit, we calculated the determination coefficient values $R^2$ for all the treatment conditions and the data from the ACT-R model. The value of $R^2 = 0.96$ for Non-White participants in \textit{Black} and \textit{White} treatment conditions and $R^2 = 0.83$ for \textit{Control} treatment. For White participants, the $R^2 = 0.93$ for \textit{White} treatment, $R^2 = 0.90$ for \textit{Control} treatment, $R^2 = 0.84$ for \textit{Black} treatment. For Black participants, $R^2 < 0.5$. Therefore, the model was accurately able to predict participant strategies for Non-White participants in \textit{Black} and \textit{White} treatment conditions.

\section{Discussion}

The statistical analysis from the experiment clearly shows how different races perceive racialized AI agents differently, so a one-size-fits-all strategy may not work as expected when deploying AI agents in real-time scenarios \cite{kim2023a}. Black participants viewed the AI agent as intelligent despite scoring lower on trials, indicating they believed that the AI agent cooperated with them. This might indicate that they believed the AI agent might have fewer biases than humans and thus could be less discriminatory \cite{kim2023a}. White and Non-White participants employed rational thinking to categorize AI agents as intelligent based on their scores, as exhibited by the weak correlation between average scores and the perceived intelligence estimate of the AI agent shown in Fig. \ref{f4}.

Qualitative analysis of survey data showed that White participants did not think AI cooperated as much in the \textit{Control} treatment condition, and Non-White participants also had similar opinions in \textit{Black} treatment conditions, but favored AI in \textit{White} treatment conditions. Eyssel et al. \cite{eyssel2013a} also found similar results where White participants displayed conditional bias towards Black-racialized robots, wherein they attributed different perceived agency for robots racialized as \textit{Black} vs \textit{White}. They also noted that social desirability may be a factor, due to which participants might not explicitly admit their racist beliefs or prejudice.

Participants of all demographics viewed the AI agent racialized as \textit{White} to cooperate the most and least work against them. This supports arguments by Cave and Dihal \cite{cave2020whiteness} that AI, from its inception, has been predominantly portrayed with White visual cues.

The strategies used by the ACT-R model scored closest to those of Non-White participants in \textit{Black} and \textit{White} treatment conditions. The model is relatively closer to the scores obtained by White participants. Decreasing base-level chunk activation of the rotation strategy, and considering the AI agent’s cooperation, has helped achieve a closer fit to participant strategies. However, the model's scores are farthest from those of Black participants.

\section{Limitations and Future Directions}

The ACT-R model we developed obtained a closer fit for Non White participants in \textit{Black} and \textit{White} treatment conditions. It can further be run with various other parameter combinations to get accurate fits for the other treatment conditions. Since the scores do not have huge variations on grouped treatments, future analysis should focus on individual participant-wise and treatment-wise behaviors to gain a complete understanding of the strategies used for decision-making and the factors impacting them.

Model parameters can be modified in certain ways to achieve the desired fit for Black participants and better understand their strategies. For example, the explicitly negative attitude towards racialized AI can be achieved by modifying the ACT-R model declarative memory-based parameters for the chunks, particularly those which are related to the strategy of observing AI's movement before deciding the next action. Similarly, other parameter settings like the procedural-based reinforcement learning parameter \textit{:alpha} for the utility learning function may be set to values other than $0.02$ to get a more accurate fit for the data collected from each treatment condition for each demographic. This can be used to test if the learning rate impacted participant scores. Another strategy that only $79$ of $935$ participants displayed during the task is the rotation strategy, so the base level activation for the declarative memory that leads to using the rotation strategy could be reduced even further given the number of participants who used this strategy for the respective treatment conditions to achieve even better individual fit. 

Instance-based learning \cite{gonzalez2003instance} can also be used to analyze individual participant strategies. Additionally, Holographic declarative memory (HDM) \cite{kelly2020a} can be used for the ACT-R model to model potential latent associations between racialized knowledge, the ways the AI agent may be racialized, and choices of strategies. Thus, integrating such memory systems for use in this model could be useful for observing the effects of sociocultural structures (tied to racialization) on decision making during this task.
Finally, we may be able to better understand the effects of racialized attitudes on behavior during the task using tasks like the Implicit Association test \cite{greenwald1998a}. 

\section{Conclusion}

In this paper, we analyzed the scores obtained by participants and the responses to the post-game survey to understand their interaction with racialized AI agents. Black participants had positive opinions on AI in all treatment conditions, indicating that race may not have impacted their behavior. However, they had notably lower scores in the \textit{Control} treatment condition, indicating that the cooperation with AI could have been triggered by racial cues attributed to the AI agent. On the other hand, White participants performed better in all treatment conditions but believed AI worked against them in the \textit{control} treatment condition based on survey responses. The negative reaction towards the AI agent, represented by lower scores or survey responses, was shown by both Black and White participants only in the \textit{Control} condition. This hints at the implicit biases they may have or their conscious effort not to report negative feelings in some cases, as shown by Kubota \cite{kubota2024a}, Eyssel and Loughnan \cite{eyssel2013a}. Non-White participants explicitly favored AI agents in the \textit{White} treatment conditions over the \textit{Black} treatment conditions, as demonstrated by their survey responses and scores. Finally, from responses to the post-game survey, we notice that all demographics favored AI agents racialized as \textit{White}. These results indicate both the implicit and explicit impact of race on the scores of participants at various degrees. 

\bibliographystyle{IEEEtran}
\bibliography{ref}

\end{document}